\documentclass[10pt,conference]{IEEEtran}
\IEEEoverridecommandlockouts
\usepackage{cite}
\usepackage{amsmath,amssymb,amsfonts}
\usepackage{algorithmic}
\usepackage{graphicx}
\usepackage{textcomp}
\usepackage{balance}
\usepackage{xcolor}
\usepackage{hyperref}
\def\BibTeX{{\rm B\kern-.05em{\sc i\kern-.025em b}\kern-.08em
    T\kern-.1667em\lower.7ex\hbox{E}\kern-.125emX}}
\begin{document}

\title{Understanding Emotions of Developer Community Towards Software Documentation}

\author{\IEEEauthorblockN{Akhila Sri Manasa Venigalla and Sridhar Chimalakonda}
\IEEEauthorblockA{\textit{Research in Intelligent Software \& Human Analytics (RISHA) Lab}\\
\textit{Department of Computer Science and Engineering} \\
\textit{Indian Institute of Technology Tirupati}\\
Tirupati, India\\
\{cs19d504, ch\}@iittp.ac.in }
}
\maketitle

\begin{abstract}

The availability of open-source projects facilitates developers to contribute and collaborate on a wide range of projects.
As a result, the developer community contributing to such open-source projects is also increasing. Many of the projects involve frequent updates and extensive reuses. A well-updated documentation helps in a better understanding of the software project and also facilitates efficient contribution and reuse. Though software documentation plays an important role in the development and maintenance of software, it also suffers from various issues that include insufficiency, inconsistency, ill-maintainability, and so on. Exploring the perception of developers towards documentation could help in understanding the reasons behind prevalent issues in software documentation. It could further aid in deciding on training that could be given to the developer community towards building more sustainable projects for society. Analyzing sentiments of contributors to a project could provide insights on understanding developer perceptions. Hence, as the first step towards this direction, we analyze sentiments of commit messages specific to the documentation of a software project. To this end, we considered the commit history of 998 GitHub projects from the GHTorrent dataset and identified 10,996 commits that correspond to the documentation of repositories. Further, we apply sentiment analysis techniques to obtain insights on the type of sentiment being expressed in commit messages of the selected commits. We observe that around 45\% of the identified commit messages express \textit{trust} emotion. 
\end{abstract}


\section{Introduction}
With developer communities growing in size,
there is a vast increase in the scale and variety of software projects. The availability of open-source platforms has facilitated easy collaborations over projects among developers all over the world \cite{dabbish2012social}. GitHub\footnote{\url{https://github.com/}} has about 40 million developers who contribute to a varied number of projects, with around 44 million repositories created during the year 2019\footnote{\url{https://octoverse.github.com/}}. Developers on GitHub contribute to projects through issue reports, pull requests, commits, and so on, and reuse the projects through forks. 
The growth of projects and developer communities has also resulted in frequent updations to the projects \cite{dabbish2012social}. The extent of contributions to and reuse of projects is influenced by the level of developer understanding about a project \cite{borges2018s}. Repositories on GitHub are accompanied by documents such as \textit{ReadMe}, and \textit{License} files, which provide information about the usage, installation, contribution guidelines, license information of a project, and so on \cite{prana2019categorizing}. 
Software documentation aids better project comprehension and plays a major role in improving the popularity of the repository and also in increasing contributions to the repository \cite{borges2016understanding}.

Software documentation is capable of aiding various phases of software development, and maintenance \cite{lethbridge2003software}. It exists in various forms such as test case documentation, API documentation, use-case documentation, and so on, to support a wide range of target audience such as testing teams, developers, and users of the system \cite{lethbridge2003software}. Software documentation is the most recommended practice for a software project \cite{aghajani2019software}. A comparison of contributions to projects on GitHub revealed that projects with frequently updated and well-written documentation have more number of contributions and were observed to be more popular \cite{borges2016understanding}. However, researchers observed several issues to be prevalent in the documentation of many projects. These issues include documentation being insufficient,  inconsistent with the project updates, and incorrect \cite{aghajani2019software}. Understanding the reason behind such issues being prevalent, in spite of the widely accepted notion of the importance of documentation,  from developers' perspective could help in taking measures to address the issues and, consequently, help in improving documentation. Existing studies in the literature to understand developer opinions towards documentation focused on developer-interviews, questionnaires and case-studies of actions during project development \cite{garousi2015usage,uddin2015api}. Emotion analysis of developer messages related to documentation in projects could also help in understanding developers' intentions towards documentation. 
Sentiments of commit messages in GitHub repositories have been analyzed to understand developer perception towards multiple phases in software development such as code refactoring, bugs in software, and so on \cite{singh2017code, sinha2016analyzing, huq2020developer}. A similar approach towards analyzing messages in developer commits, specific to documentation files in GitHub projects, could help in obtaining useful insights about developer perception towards documentation in the project. 
The existing studies in literature, performed through questionnaires and interviews of developers, reveal that developers consider documentation to be useful but do not update documentation frequently, which might imply a negative perception towards updating documentation \cite{aghajani2019software, garousi2015usage}. Based on the discussions in the literature about issues in software documentation and the observation of the presence of more negative emotions in commit messages \cite{sinha2016analyzing}, we start our exploration of emotions in documentation-related commit messages on projects in GitHub, with the following hypothesis.

\textit{\textbf{Hypothesis:} \textbf{Developers have a negative perception towards updating documentation, which could be fear, anger, disgust, or sadness.}}

Based on the interviews conducted by researchers in existing literature \cite{aghajani2020software, aghajani2019software, sinha2016analyzing}, where developers have stated that they find the majority of the documentation to contain several issues, though they agree it to be an important artifact, and the negative emotions identified in commit messages, we assume the hypothesis considered to be a general perception among the developer community and researchers. 
This paper explores the commit history of 998 randomly selected projects on GitHub\textsuperscript{3}. It analyzes sentiments expressed in documentation-related commit messages, intending to identify frequently expressed emotions by developers in documentation-related commits and developer perception towards documentation in GitHub projects.

\section{Related Work}

Sentiments of developer discussions and other developer messages on GitHub have been analyzed to understand developers' perceptions towards various artifacts in the repositories \cite{pletea2014security, jurado2015sentiment, guzman2014sentiment, sinha2016analyzing}. Pletea et al. \cite{pletea2014security} analyze commit messages and pull requests corresponding to security-related discussions to understand developers' emotions towards security concerns in the software project. This analysis reiterates the importance of providing appropriate training to developers towards handling security-related issues. Sentiments of issues in nine GitHub projects were analyzed, with the aim to verify the contribution of emotions in issues towards obtaining insights on the development process of the projects \cite{jurado2015sentiment}. This case study reveals that the emotions of developers could be obtained from the text in the issues raised and could be used in assessing the development of projects. 

Guzman et al. \cite{guzman2014sentiment} have analysed sentiments of around 60K commit messages from 29 projects on GitHub,  with respect to the corresponding programming language used, day of the week, time of logging the commit message, nature of the team, and so on. More negative commits were observed to be logged for Java language and on Mondays. Another study on sentiments in commit logs of around 28K projects reveals that negative sentiments exist 10\% more than positive sentiments in commit messages \cite{sinha2016analyzing}. Further, in this study, Sinha et al. \cite{sinha2016analyzing} have also observed that more negative sentiments are exhibited on Tuesdays in comparison with any other days of the week. Sentiments of around 3K commit messages from 60 Java projects on GitHub were studied to understand the developers' emotion during code refactoring, which reveals that the majority of the developers tend to express negative emotions during the process of code refactoring \cite{singh2017code}. Another comparative study on sentiments of commit messages for regular commits and those preceding bug fixes identifies more negative emotions in the case of commits that are logged against bug fixes \cite{huq2020developer}.  

Werder et al. \cite{werder2018meme} have developed a method by the name \textit{MEME} to extract commit messages and other comments of projects from GitHub and analyze the sentiments of these texts.

Several works exist in the literature towards assessing sentiments of commit messages and issue reports to obtain insights about various artifacts and actions in the projects such as security-related aspects, code refactoring, programming languages, and so on \cite{singh2017code, sinha2016analyzing, huq2020developer, guzman2014sentiment, jurado2015sentiment}. Understanding the sentiments of developers towards documentation in a project could help understand the reasons behind the issues that are prevalent in software documentation. However, we did not find any study that tries to analyze the emotion of documentation-related commit messages. Hence, we propose a study to analyze the sentiment of documentation-related commit messages in projects on GitHub.

\section{Methodology}
\begin{table*}[]
    \caption{Example Commit Messages for each Emotion Class}
    \label{tab:commitmsg}
    \centering
    \begin{tabular}{|c|c|c|}
    \hline
   \textbf{ Repository Name} & \textbf{Commit Message} & \textbf{Emotion}  \\
    \hline
     bittorrent & \textit{\textbf{Improve} duplicate torrent detection} & Trust\\
     
     disqus-notify-content-author & \textit{README is better \textbf{short-n-sweet}} & Joy\\
     
     SynappsBundle  &  \textit{Improved test framework, added \textbf{missing} dependencies} & Fear\\
     
     WDS-Required-Plugins & \textit{Make usage \textbf{instructions} more clear} & Anticipation\\
     
     media-bundle & \textit{\textbf{drop} support for older versions} & Sadness\\
     
     yoapp-php & \textit{Location support (backward compatibility \textbf{broken})} & Anger\\

    bb-legacy-plugins & \textit{plugin-browser-for-bbpress: A \textbf{stupid} bug. Tagged version 0.1.9} & Disgust\\
    
    ccglite &  \textit{\textbf{Didn't expect} this to actually become a thing} & Surprise\\
    \hline
    \end{tabular}
\end{table*}

\subsection{Dataset}
To understand the emotions of developers towards documentation in software projects, we considered projects available on the GitHub platform. We selected 1000 projects randomly from the existing GHTorrent dataset \cite{Gousi13}. We explored the readme files of these repositories and excluded two repositories that were written in a non-English language, resulting in a total of 998 GitHub repositories. We then scraped the commit log for each of the 998 projects. We aim to assess the emotions in documentation related commit messages for the considered 998 projects. Towards this, we primarily identified the list of commits that are related to the documentation of the project. Documentation-related files were identified based on file names and file extensions in literature \cite{prana2019categorizing, aggarwal2014co,  vendome2017license}.
Aggarwal et al. \cite{aggarwal2014co} mentions the presence of documentation related files with the extensions - `.txt', `.md', `.png', `.jpg', `.mp4'. This work provides us insights on type of files that can be considered to be related to documentation. Furthermore, researchers specify existence of documentation in files containing \textit{readme} and \textit{license} as a part of their labels in the projects\cite{vendome2017license, prana2019categorizing}.       

Based on the existing literature on software documentation, textual documentation of projects was observed to be present in files that are labeled as `\textit{Readme}', `\textit{License}` and other text files. 
Considering this observation as a basis, we identified commits that modified all the files containing `\textit{ReadMe}' or `\textit{License}' as a substring in their labels, and the files that have `.txt' or `.md' as extensions.
Filtering the commits resulted in 6549 commits that updated documentation-related files in the repositories. These 6549 commits were observed to contain 10,996 commit messages, leading to a dataset of 10,996\footnote{\url{https://github.com/AkhilaSriManasa/Understanding-Emotions-of-Developer-CommunityTowards-Software-Documentation}} commit messages. We thus analyse emotions conveyed in a total of 10,996 commit messages.

\subsection{Emotion Analysis}
To analyze emotions being conveyed in the commit messages, we loaded all the identified commit comments as corpora and cleaned the text to remove special characters and numerics in the text. We then applied further transformations on the corpora to remove stop words of the English language in the text. Further, we removed any white-spaces present in the text and performed stemming on the text. This resulted in a set of words in their root forms in the text of each commit message.
After cleaning the text, we performed Emotion Analysis using \textit{Syuzhet} package. \textit{Syuzhet} package contains \textit{nrc} module, referring to 
\textit{NRC Word-Emotion Association Lexicon\footnote{\url{https://saifmohammad.com/WebPages/NRC-Emotion-Lexicon.htm}}}, that identifies presence of eight categories of emotions in a given text. Though \textit{MEME} has been proposed to analyse sentiments of artifacts in GitHub repositories, we could not find the tool openly available to perform the sentiment analysis. However, we observed that \textit{MEME} has also been developed using the \textit{nrc} module of \textit{Syuzhet} package. Hence, we assume that the results of emotion analysis would be similar to that of applying \textit{MEME} on the dataset. For a given corpora of text, the \textit{nrc} module of \textit{Syuzhet} package produces list of scores for each commit message with respect to eight emotion categories - \textit{anger, anticipation, disgust, fear, joy, sadness, surprise, trust} and for \textit{negative} and \textit{positive} emotions. The scores obtained indicate the number of instances of words that belong to each of the emotion categories. As a result, the emotion of each commit message could be identified based on the score obtained for the commit message in each of the emotion categories. 

There is a possibility of one commit message expressing more than one emotion or no emotion at all. However, we consider the most strongly expressed emotion to be the emotion of the commit message. Table \ref{tab:commitmsg} presents one example commit message for each emotion class, with the closest word representing the emotion specified as bold text. Table \ref{tab:commitmsg} also presents the corresponding GitHub repository for each commit message. For example, \textit{``plugin-browser-for-bbpress: A stupid bug. Tagged version 0.1.9"} is a commit message that was logged in \textit{``bb-legacy-plugins"} repository. The commit message was analyzed and observed to contain \textit{disgust} emotion. The presence of \textit{stupid} word in the commit message might have influenced the emotion of the message, as a result of which, the message was observed to express \textit{disgust} emotion. 
The considered commit messages might also express multiple emotions but are presented based on the strongest emotion being expressed. 

\section{Results}
\begin{figure}
    \centering
    \includegraphics[width = \linewidth]{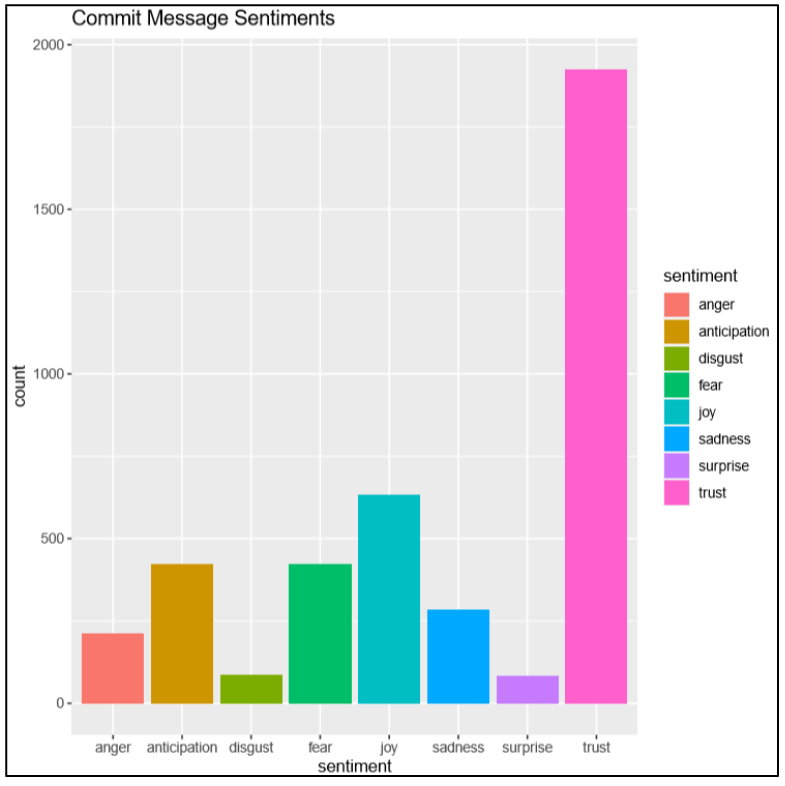}
    \caption{Results of Emotion Analysis of Commit Messages}
    \label{fig:results}
\end{figure}

Results of the emotion analysis are presented in Fig \ref{fig:results}.  
For better comprehension and to identify the most expressed emotions among all commit messages, we plot the results in a graph, with emotion categories on X-axis and number of instances of words in the corpus, corresponding to the respective emotion classes on Y-axis, as shown in Fig \ref{fig:results}.
The results indicate that majority of the words in the commit messages correspond to \textit{trust} emotion, followed by \textit{joy} and \textit{fear} emotions. We also observe that around 45\% of the text express \textit{trust} emotion, 15\%  of the text expresses \textit{joy} emotion, around 10\% of the text expresses \textit{fear}, \textit{anticipation} and \textit{sadness}, while the cumulative of rest three emotions were observed to be around 10\% of the total text. Further, it was also observed that \textit{positive} emotions exist in 78.4\% of the text and \textit{negative} emotions exist in 21.5\% of the text. 
\fbox{\begin{minipage}{25em}
Out of 10,996 comments analysed, we observed that more than 4948 comments expressed  \textit{trust} emotion, while only 220 comments expressed \textit{disgust} emotion, thus, negating our hypothesis.
\end{minipage}}
\newline

The most frequent emotion category being \textit{trust} might imply that most of the developers have trust in the existing documentation and mostly rely on the existing documentation. 
This would also imply that the updates to documentation related files are logged confidently and believed to be correct by the collaborators, reducing the scope of further discussions on modifying the information in these files. These minimal discussions about improving the documentation-related files could be one of the reasons for incomplete or inconsistent documentation.  

The results also indicate that positive emotions are greater than negative emotions in documentation-related commit messages. This revelation of frequent emotion category and the greater number of positive emotion contradicts our hypothesis that developer-community has a negative perception such as \textit{fear, anger or sadness} towards software documentation. Contradicting our hypothesis, \textit{fear, anger and sadness} were observed to exist in around 10\% of the text each, succeeding \textit{trust} and \textit{joy}. Comparing the obtained results against the widely accepted notion of documentation having issues such as incompleteness, inconsistency, and so on, we could deduce that developers might only make minor changes in the existing documentation due to increased trust. This positive emotion towards documentation among developers could be one of the factors for deficient documentation, as the developers might be trusting the existing documentation. However, a clear understanding of the prominent \textit{trust} emotion observed in the results could be obtained by a manual walk-through of the commit messages that correspond to \textit{trust} emotion.

\section{Threats to Validity}
The emotion analysis of developer community is performed solely on commit messages in this paper. However, developers perform multiple actions on GitHub, such as logging issues, performing pull requests, and so on. These actions also include textual information in the form of title and comments appended with the artifacts, i.e., issues and pull requests. Considering artifacts other than commits, such as issues, developer discussions, pull requests, and so on, might yield varied results for the considered set of projects.

The commit messages considered for analysis are based on the assumption that all files with `.txt' and  `.md'  as extensions and files with `readme' or `license' in their labels are documentation-related files. There might exist some other files which do not satisfy the specified criteria but still be related to documentation of the project. The relation of these files with documentation might have to be manually analyzed by inspecting content in the files. However, while considering numerous projects, it is difficult to manually review each file in the project and identify its relationship with documentation. The current format for existing commit messages does not facilitate identifying its relation to the documentation of the project, apart from string matching with the file names. Also, \textit{init}, \textit{readme}, \textit{version}, \textit{add} and \textit{create} were observed to be the top 5 frequently occurring words in commit messages. It is difficult to deduce any emotion based on any of these five words, indicating that the current format of commit messages does not explicitly convey developers' intent.

Another main limitation of this analysis is that the commit messages are extracted from commit logs. There might exist commits that modify multiple files, along with documentation-related files in a project. Such commits might contain only one message for all modified files, which might not be specific to documentation-related files. Other approaches in Machine Learning and Natural Language Processing, to identify commit messages that specifically belong to only documentation files, could be explored in the future versions of this work.

The accuracy of emotion analysis is largely dependant on the \textit{nrc} module of \textit{syuzhet} package. Varying the sentiment analyzer might produce different results. Other emotion analyzers and better approaches towards analyzing emotions of commit messages could further be explored in future versions. Also, the obtained results are specific to the 998 projects considered and might vary for a different set of projects.

\section{Conclusion and Future Work}
Considering the importance of documentation and the existent issues in documentation, we aimed to understand the perception of the developer-community towards software documentation. Towards this aim, we considered documentation-related commit messages of 998 GitHub projects, randomly selected from the GHTorrent dataset \cite{Gousi13}. 
We extracted documentation-related commit messages based on criteria derived from existing literature for files corresponding to the commits from the 998 projects. This resulted in 6549 commits, comprising a cumulative of 10,996 commit messages.

We started our experiment with the hypothesis that the developer-community might have a negative perception towards documentation. However, the emotion analysis results contradict our hypothesis and reveal that developers express more \textit{positive} emotions than \textit{negative} emotions in documentation-related commit messages. It has also been observed that among all the 8 emotions, \textit{trust} is the most frequently expressed emotion in commit messages.  Thus, we observe that \textit{trust} is the most frequently expressed emotion by the developer community in documentation-related commits and that developers might perceive the existing documentation to be reliable and trust-worthy.

As a part of the future work, we plan to analyze other artifacts in software projects, such as issue comments, pull request comments, developer discussions, and so on. We further plan to manually identify the emotion of a few commit messages and then assess the accuracy of \textit{nrc} emotion analyzer module of \textit{Syuzhet} package by comparing the results of emotion analysis with manual labeling. We also plan to implement multiple emotion analysis tools to identify the most accurate emotion analyzer based on a comparison of emotion categories identified, with manual labeling.

\section{Data Availability}
The dataset extracted for this study and the corresponding source code used to perform the study is presented at - \url{https://doi.org/10.5281/zenodo.4569463}.

\balance
\bibliography{references}
\bibliographystyle{IEEEtran}

\end{document}